\documentclass[12pt,letter]{article}

\usepackage[dvipsnames]{xcolor}
\usepackage[most]{tcolorbox}
\usepackage{geometry}
\usepackage{color,xcolor}
\usepackage[space]{grffile}
\usepackage{latexsym}
\usepackage{textcomp}
\usepackage{longtable}
\usepackage{multirow,booktabs}
\usepackage{amsfonts,amsmath,amssymb}
\usepackage{url}
\usepackage[english]{babel}
\usepackage{graphicx}
\usepackage[space]{grffile}
\usepackage{url}
\usepackage[utf8]{inputenc}
\usepackage{longtable}
\usepackage[sort, numbers]{natbib}
\usepackage{ulem}
\usepackage{fancyhdr}
\usepackage{authblk}
\usepackage{hyperref}
\hypersetup{
   colorlinks,%
   citecolor=blue,%
   filecolor=black,%
   linkcolor=blue,%
   urlcolor=blue
}

\newcommand{\msun}{\,{\rm M}_{\odot}}

\definecolor{arylideyellow}{rgb}{0.91, 0.84, 0.42}
\definecolor{corn}{rgb}{0.98, 0.93, 0.36}
\definecolor{pastelyellow}{rgb}{0.99, 0.99, 0.59}

\title
{\large {Astro2020 Science White Paper:  THE GRAVITATIONAL WAVE VIEW OF MASSIVE BLACK HOLES}}

\author{\small M. Colpi$^1$\footnote{monica.colpi@mib.infn.it}, K. Holley-Bockelmann$^{2,3}$\footnote{k.holley@vanderbilt.edu}, T. Bogdanovi\'c$^4$, P. Natarajan$^5$, J. Bellovary$^{6}$, A. Sesana$^{1,7}$, M. Tremmel$^8$, J. Schnittman$^{9}$, J. Comerford$^{10}$, E. Barausse$^{11}$, E. Berti$^{12}$, M. Volonteri$^{13}$, F. M. Khan$^{14,2}$, S. T. McWilliams$^{15}$, S. Burke-Spolaor$^{16}$, J. S. Hazboun$^{17}$, J. Conklin$^{18}$, G. Mueller$^{19}$, S. Larson$^{20}$}

\date{}

\begin{document}
\maketitle

\vspace{-32pt}
\begin{figure}[h]
\begin{center}
\includegraphics[trim=35mm     
65mm 30mm 75mm, clip, width=1.000\textwidth]{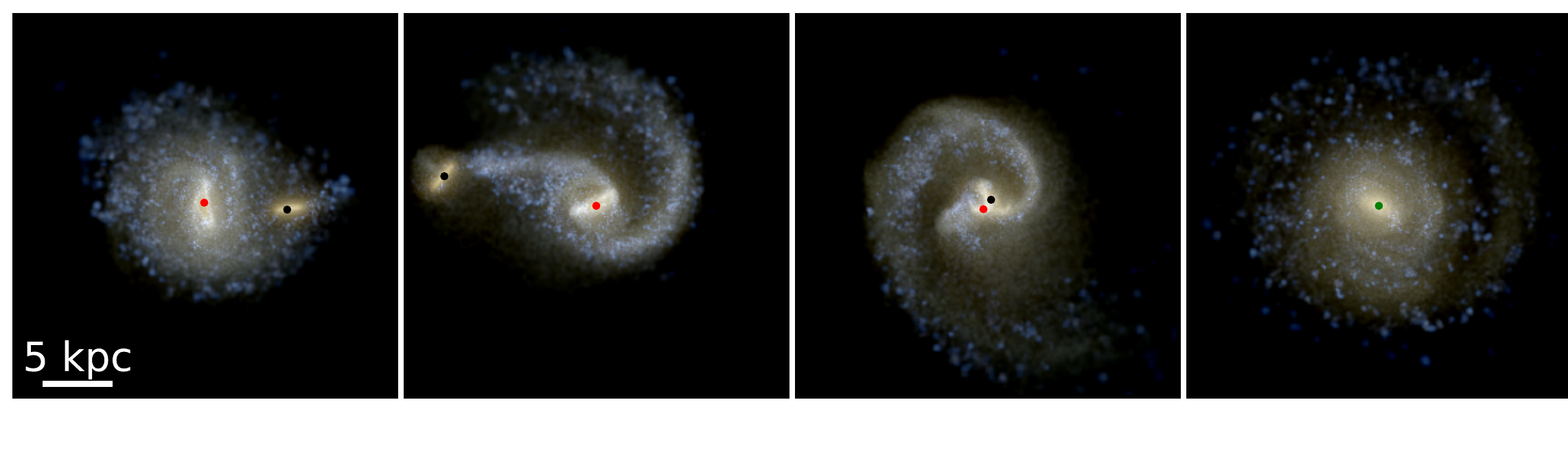}

%\caption{Figure 1 reproduced with permission from \citet{tremmel18}}
\label{front} 
\end{center} 
\end{figure}
\vspace{-58pt}
\begin{center}
{\scriptsize Figure depicting the merger of two galaxies with their nuclear MBHs (circles), adopted with permission from \citet{tremmel18}}
\end{center}
\vspace{-12pt}
 
\noindent\fcolorbox{white}{white}{%
    \minipage[t]{\dimexpr1.00\linewidth-2\fboxsep-2\fboxrule\relax}
%\begin{center} {\large\bf Summary} \end{center}
{\small Coalescing, massive black-hole (MBH) binaries are the most powerful sources of gravitational waves (GWs) in the Universe, which makes MBH science a prime focus for ongoing and upcoming GW observatories. The Laser Interferometer Space Antenna (LISA) -- a gigameter scale space-based GW observatory -- will grant us access to an immense cosmological volume, revealing MBHs merging when the first cosmic structures assembled in the Dark Ages. LISA will unveil the yet unknown origin of the first quasars, and detect the teeming population of MBHs of $10^{4-7}\msun$ 
forming within protogalactic halos. The Pulsar Timing Array, a galactic-scale GW survey, can access the largest MBHs the Universe, detecting the cosmic GW foreground from inspiraling MBH binaries of $\sim 10^9\msun$. LISA can measure MBH spins and masses with precision far exceeding that from electromagnetic (EM) probes, and together, both GW observatories will provide the first full census of binary MBHs, and their orbital dynamics, across cosmic time. 
%and uncover their merger and accretion history across cosmic time, in a unique way. 
Detecting the loud gravitational signal of these MBH binaries will also trigger alerts for EM counterpart searches, from decades (PTAs) to hours (LISA) prior to the final merger. 
By witnessing both the GW and EM signals of MBH mergers, precious information will be gathered about the rich and complex environment in the aftermath of a galaxy collision.
%LISA observations in a different mass range and at lower redshifts. 
The unique GW characterization of MBHs will shed light on the deep link between MBHs of $10^4-10^{10}\msun$ and the grand design of galaxy assembly, as well as on the complex dynamics that drive MBHs to coalescence.
}
%What is the massive black hole coalescence rate and how is it related to the galaxy merger rate? 
%What can the spin of coalescing massive binary black holes tell us about their evolution ruled by mergers and accretion? 
     \endminipage}\hfill\\

\noindent
\begin{center}

{\bf Thematic Science Areas:} Galaxy Evolution, Multi Messenger  Astronomy  and  Astrophysics, Cosmology  and  Fundamental  Physics 
\end{center}

\newpage

%%%%%%%%%%%%%%%%%%%%%%%%%%%%%%%%%%%%%%%%%%%%%%%%%%%%%%%%%%%%%%
%% Section 1
%%%%%%%%%%%%%%%%%%%%%%%%%%%%%%%%%%%%%%%%%%%%%%%%%%%%%%%%%%%%%%
\noindent{\Large \bf A New Window into the Cosmos} \label{sec:intro}
%\vspace{-3mm}
\medskip

\noindent Gravity has its own messenger: GWs are ripples in the fabric of spacetime produced by non-axisymmetric motions of matter. Traveling essentially unimpeded throughout the Universe, GWs carry unbiased information on their sources, from binary stellar remnants, to MBH collisions, to the Big Bang itself. GWs provide a clean way to measure the geometry of black hole spacetimes, including masses and spins, and even characterize their horizons \citep{Klein16,Berti2016,Cardoso2017}. With their strongly curved geometry and relativistic motion, coalescing MBH binaries generate a highly warped and dynamic spacetime -- the strongest gravitational signals expected in the Universe. Moreover, their amplitude and frequency show a simple and universal scaling with mass, inherited from the fact that general relativity has no built-in fundamental scale. This gives us direct access to a huge range of black holes -- from primordial to ultra-massive -- by exploring different GW bands. With gravity as a messenger, we stand to revolutionize our understanding of the birth, growth, and evolution of MBHs, as well as their role in sketching the cosmological canvas of the Universe. 

\medskip

\noindent%\fcolorbox{white}{white}{%
\minipage[!t]{\dimexpr0.55\linewidth\fboxsep\fboxrule\relax}
    \noindent Most of what we know about MBHs thus far has been informed by EM observations of active galactic nuclei (AGN) over several epochs of cosmic evolution \citep{heckman14,madau14}. During the {\sl Cosmic Dawn}, starting at $z\sim 20$, mostly-neutral baryons in low-mass dark matter halos began to collapse and fragment, forming the first stars and seed black holes. The physics in this era is all but invisible to us in the EM window until it ends at $z=7.5$, when a myriad of sources of ultra-violet radiation, including accreting MBHs, reionized the intergalactic neutral hydrogen into a hot, tenuous plasma \citep{Planck-2018-cosmology}. The most
\endminipage \hfill
%\noindent\fcolorbox{black}{pastelyellow}{%
    \minipage[!t]{\dimexpr0.42\linewidth\fboxsep\fboxrule\relax}
%    \vspace{12pt}

\begin{tcolorbox}[colback=pastelyellow,%gray background
                  colframe=black,% black frame colour
                  width=7cm,% Use 5cm total width,
                  arc=3mm, auto outer arc,
                 ]
 
%\end{tcolorbox}
\vspace{5pt}
\begin{center} {\bf MBH Mysteries} \end{center}
{ 
%$\bullet$ What seeded the formation of the rare, bright quasars shining at redshifts as large as $z\sim 7$?\\
%$\bullet$ How did these giant BHs grew massive in less than a billion of years after the Big Bang?\\

$\bullet$ How are MBHs born and how do they grow?\\
\vspace{-2.0mm}

\noindent$\bullet$ How efficiently do MBHs merge and how does this affect their galaxy hosts?\\
\vspace{-2.0mm}

$\bullet$ What are the demographics of MBHs in the Universe?\\
%in terms of their masses, spins and binarity?
%in terms of their masses, spins and binarity?
\vspace{-4.0mm}

%$\bullet$ What is the redshift at which the earliest binary BHs appear, and what is their mass and spin?\\
%$\bullet$ What does their redshift tell us about their formation pathway and dynamics of pairing?\\$\bullet$ Do mergers play a key role in the early build up of massive BHs?\\
%$\bullet$ What does spins tell us about BH evolution ruled by mergers and accretion?\\
%$\bullet$  Are the high-$z$ binary BHs of $10^{4-5}\msun$ the seeds upon which the giant BHs have grown? \\
%$\bullet$ What is the massive BH coalescence rate and how does it relate to the galaxy merger rate?
}

\end{tcolorbox}
\endminipage\hfill
 
\noindent distant quasars are now found at $z\sim 7$, when the Universe was less than one billion years old, posing extreme constraints on their formation squarely in this heretofore unobserved era\citep{Banados18}. 
    These rare, overluminous sources are probing the tip of an underlying population of yet undiscovered much fainter objects. GW observations will be key to unveiling the existence of and physics governing MBHs within the Cosmic Dawn.

Well after MBH seeds are sown comes the epoch of{\sl Cosmic Noon}, extending from $z\sim 6$ to $2$. This is the epoch of galaxy growth through repeated major mergers, accretion of lower mass dark-matter halos, and cold gas flowing in along dark matter filaments \citep{white78,barnes96,white91,dekel09}.
Around $z\sim 2$, the cosmic integrated star formation rate and AGN activity reach their peak, followed by a decline that extends to the present day \citep{heckman14}. Though it is widely accepted that AGN activity and major mergers are related, the precise details of how MBHs assemble during this epoch is an open question, one that GWs provide unique data to answer.

%%%%%%%%%%%%%%%%%%%%%%%%%%%%%%%%%%%%%%%%%
%  Box 1
%%%%%%%%%%%%%%%%%%%%%%%%%%%%%%%%%%%%%%%%%
\smallskip
%\noindent\fcolorbox{white}{white}{%
%    \minipage[!t]{\dimexpr0.58\linewidth-2\fboxsep-2\fboxrule\relax}
Upcoming GW observations serve complementary views of the cosmos. GWs at low frequency will observe the origin and evolution of the most extreme and enigmatic objects in the Universe:  {\it massive black holes}. In this whitepaper, we describe the science that can be obtained by LISA, the first-generation space-based GW observatory. By design, LISA will provide key observations needed revolutionize out view of MBHs \citep{LISA17}, filling an unobserved gap of $\sim$9 orders of magnitude between nHz, where Pulsar Timing Arrays (PTAs) are sensitive to supermassive black holes orbiting on timescales of decades \citep{IPTA16}, and $>$Hz, where ground-based observatories probe the last fraction of a second of stellar mass black hole mergers.

\bigskip

%%%%%%%%%%%%%%%%%%%%%%%%%%%%%%%%%%%%%%%%%%%%%%%%%%%%%%%%%%%%%%
%% Section 2
%%%%%%%%%%%%%%%%%%%%%%%%%%%%%%%%%%%%%%%%%%%%%%%%%%%%%%%%%%%%%%
\noindent {\Large \bf Massive Black Holes in the Gravitational Universe}

\medskip

The GW signals from comparable mass ratio coalescing MBHs are similar in shape to the first signal ever detected, GW150914 \citep{Abbott-1}, and the subsequent stellar black-hole binary mergers. Indeed, one simply needs to appropriately rescale the time variable -- making the signal longer lasting (from seconds to months)-- and the amplitude -- which results in signal-to-noise ratios (SNRs) as high as $\sim 1000$, to be compared to SNRs of a few tens at most for today's Advanced LIGO and Advanced Virgo. Because of this rescaling, the GW frequency of merging binary MBHs with total masses of $ 10^4\msun - 10^7\msun$ falls squarely within LISA's bandwidth (which extends from about 100 $\mu$Hz to 100 mHz) in the late inspiral, merger and ringdown phase of the binary evolution. The best sensitivity will be
 reached for MBHs with masses comparable to the one residing at the heart of the Milky Way, i.e. $\sim 10^5\msun -10^6\msun$.  
 The galaxy mass function suggests that these `low-mass' MBHs are the most common, but the least well-known in terms of basic demographics, birth, growth, dynamics and connection to their galaxy host \citep{Kormendy13}.
 %during which the two black holes can still be considered as structureless is visible as a chirp. It is then followed by the ,
%i.e. the non linear plunge-in phase and coalescence, when the two horizons touch to form a new black hole, and by the {\it ring-down}, the phase describing the rapid settling of the new  black hole whose spacetime is vibrating, emitting all residual asymmetries.
%The signal  consists of a superposition of quasi-normal modes whose frequencies and decaying timescales  depend
%only on the mass and spin of the new black hole, according to the "no-hair" theorem of general relativity.  
%GWs carry away linear momentum when the binary has  %\cite{Lousto11}.    unequal masses, or spins, or a combination of the two. The new black hole receives a gravitational recoil of up to $\sim 4000\,\kms$, for specific spin configurations.

%%%%%%%%%%%%%%%%%%%%%%%%%%%%%%%%%%%%%%%%%
%  Box 2
%%%%%%%%%%%%%%%%%%%%%%%%%%%%%%%%%%%%%%%%%
\smallskip
%\noindent\fcolorbox{white}{white}{%
%    \minipage[!t]{\dimexpr0.50\linewidth-2\fboxsep-2\fboxrule\relax}    
Figure 1 shows the vastness of the LISA exploration volume: lines of constant SNR are depicted in the $M_{\rm B} - z$ plane, where $M_{\rm B}$ is the mass of the binary in the source frame. LISA will be unique at detecting the GW signal from coalescing binaries between $10^4\msun$ and a few $10^7\msun$, with SNR higher than 20 at formation redshifts $z$ as 
large as 20, and SNR as high as a few thousands at low redshifts~\citep{Klein16,LISA17}.
\bigskip
%LISA will enable the  measurement of the source frame masses and luminosity distance with fractional error of 20\% to distinguish among the seed formation channels. 
% \endminipage}\hfill

\noindent \begin{minipage}{9cm}
 \begin{flushleft}
\noindent \begin{tcolorbox}[colback=pastelyellow,%gray background
                  colframe=black,% black frame colour
                  width=8.8cm,% Use 5cm total width,
                  arc=3mm, auto outer arc,
                 ]
 
%\end{tcolorbox}
\vspace{5pt}
%\noindent\fcolorbox{black}{pastelyellow}{%
%    \minipage[!t]{\dimexpr0.47\linewidth-2\fboxsep-2\fboxrule\relax}
%    \smallskip
\begin{center} {\bf Solving MBH Mysteries} \end{center}

%$\bullet$ What seeded the formation of the rare, bright quasars shining at redshifts as large as $z\sim 7$?\\
%$\bullet$ How did these giant BHs grew massive in less than a billion of years after the Big Bang?\\

$\bullet$ LISA will measure the masses and spins of coalescing MBHs to a few $\%$ accuracy in $10^4 - 10^7\msun $ binaries out to $z\sim 20.$\\
%, and luminosity distances with uncertainties up to 23 \% accuracy, limited by weak lensing.\\ 
%tracing this population from its inception to the present day.\\
\vspace{-1.5mm}

$\bullet$ GWs will unveil the MBH growth via mergers, and their accretion history via mass and spin measurements.\\
\vspace{-1.5mm}

$\bullet$ GWs will shed light on the co-evolution of galaxies and MBHs.\\
\end{tcolorbox}  \end{flushleft} \end{minipage} \hspace{0.4cm} \begin{minipage}{6.75cm}% create a content half width
%   \minipage[!t]{\dimexpr0.50\linewidth-2\fboxsep-2\fboxrule\relax}
%\begin{flushright}
Detecting coalescing MBH binaries with $M_{\rm B}$ of $10^4-10^6\msun$ at $z>10$ will provide unique insights into the initial masses, occupation fraction, and early growth of the first seed BHs, ancestors of the MBHs \citep{Volonteri10,Natarajan14}. This will inform us about the physics producing the seeds, whether it involves the first generation of metal-free stars, the direct collapse of massive clouds \citep{Latif16}, the collapse of hyper-massive stars formed in stellar runaway collisions \citep{Devecchi2012},  or a different process altogether, such as primordial black holes formed in the early Universe, before the epoch of galaxy formation~\citep{Bernal2018}. Knowledge 
%\end{flushright}
\end{minipage}

%\noindent in the early Universe, before the epoch of galaxy formation~\citep{Bernal2018}. 
%The limit to LISA's accuracy in measuring the luminosity distance (about 20\%) 
%is due to weak gravitational lensing \citep{Holtz05}. 
\noindent of this population will anchor the initial conditions of MBH cosmic evolution, setting the stage for their subsequent mass growth and merger rate. 
%\end{flushright}
%\end{minipage}
%}

\begin{figure}[t]
\begin{center}
%\noindent\fcolorbox{white}{white}{%
%    \minipage[!t]{\dimexpr0.7\linewidth-2\fboxsep-2\fboxrule\relax}
\includegraphics[width=15.5cm]{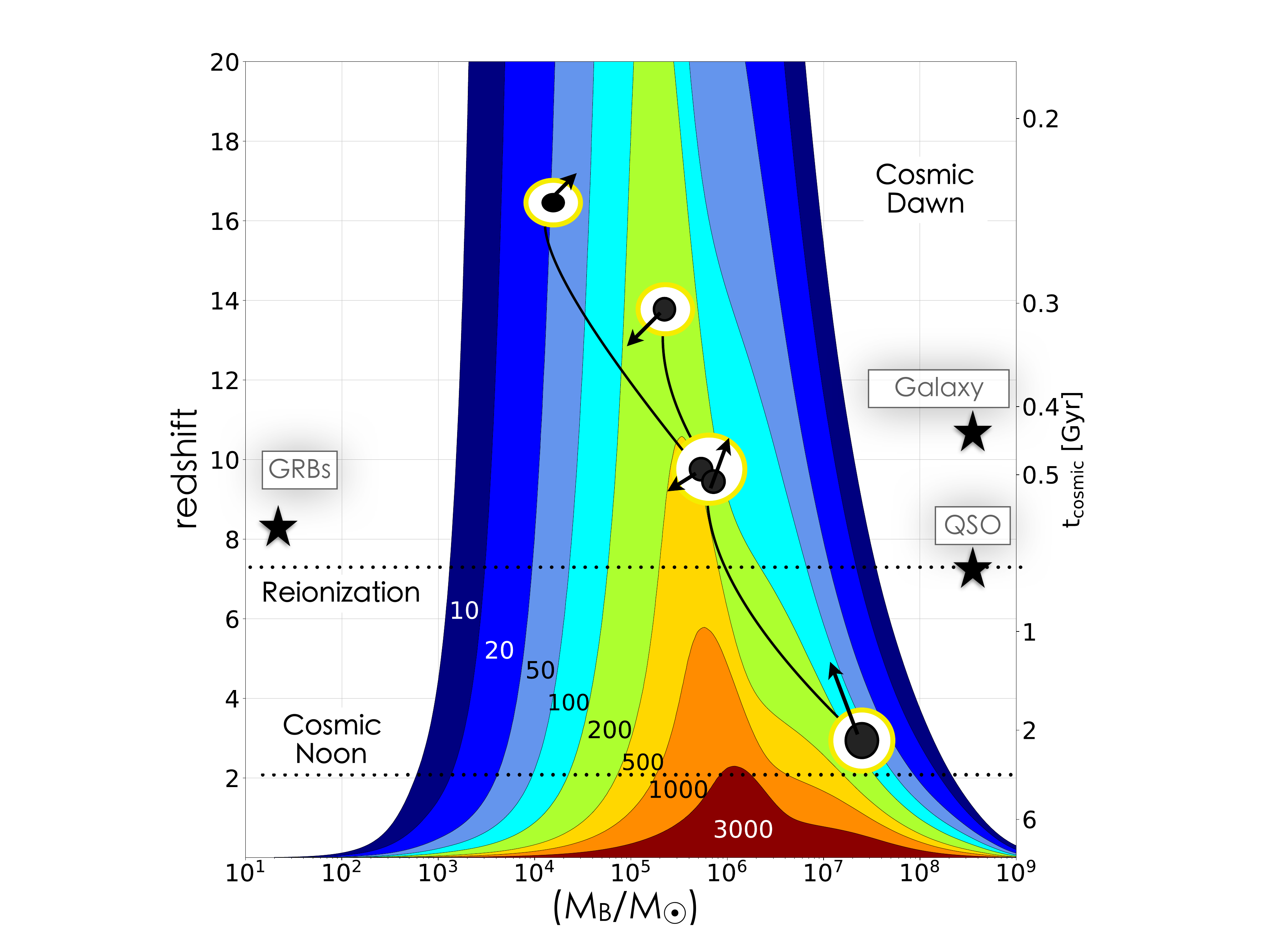}
% \endminipage}\hfill
%   \noindent\fcolorbox{white}{white}{%
%  \minipage[!t]{\dimexpr0.3\linewidth-2\fboxsep-2\fboxrule\relax}
\caption{\small Contours of constant SNR as a function of redshift (cosmic time) and source-frame binary mass $M_{\rm B}$ for the LISA observatory. For this figure, the MBHs are non spinning and have mass ratio $q=0.5$. Overlaid is an illustration of evolutionary tracks ending with the formation of a MBH at $z\sim 3$. Black dots and arrows represent the MBHs and their spins, respectively. MBHs are embedded in galaxy halos (white-yellow circles) and experience episodes of accretion (black lines) and mergers. Black stars refer to the most distant long Gamma-Ray Burst host, quasar and galaxy detected so far.}
\end{center}
\label{waterfall-LISA} 
\end{figure}

  Between about $3\lesssim z\lesssim 10$, LISA will detect the inspiral, merger and ringdown of sources with $10^5<M_{\rm B}/\msun<10^7,$ enabling the measurement of their intrinsic masses with accuracies at the 
  %3\% 
  percent level \citep{Klein16}. GW signals will also carry exquisite information on the MBH spins, which enhance the GW amplitude, introducing modulations in the signal due to precession. Spins of MBHs powering AGN are difficult to measure from their EM spectra \cite{Reynolds14}.
  %, and often their orientation cannot be determined. 
  On the contrary, the spin of the larger (smaller) MBH in a binary will be measured from the GW signal with absolute error of better than 0.01 (0.1) in several LISA events, and the spin misalignment relative to the orbital angular momentum will be determined to within 10 degrees or better~\citep{Klein16}.  Individual spins prior to the merger encode information  on whether accretion, which shaped both the MBH mass and spin evolution, was coherent (leading to spins close to maximal and small misalignment angles) or chaotic (leading to lower average spins and random spin orientations over the black hole life cycle)~\citep{King05,Barausse2012,Sesana14spin}. This will give us the unprecedented opportunity to reconstruct the MBH cosmic history from GW observations alone \citep{BertiVolonteri08}.  Moreover, 
  by measuring the angle between BH spins and the angular momentum, crucial information will be gathered about 
  the interaction between the MBHs and their environment and about whether the binary evolution is driven by the gas. In particular, gas can exert dissipative torques on the BH spins, potentially aligning them with the gas angular momentum. This also has crucial implications for the fate of the MBH produced by the merger, which could be imparted a large GW recoil velocity in the presence of large spin orbit misalignment prior to merger  \citep{Campanelli07,Bogdanovic07,Haiman09,Dotti10,Kesden2010,Roedig12,Berti2012,Lousto12,Miller13}.

%%%%%%%%%%%%%%%%%%%%%%%%%%%%%%%%%%%%%%%%%%%%%%%%%%%%
%  FIGURE 1
 %%%%%%%%%%%%%%%%%%%%%%%%%%%%%%%%%%%%%%%%%%%%%%%%%%%%

%%%%%%%%%%%%%%
 As illustrated in Figure 1, GWs from  MBHs are incredibly strong, and the advantage of this fact cannot be understated. At Cosmic Noon, right when galaxy mergers are rife, MBH mergers become extraordinarily loud, which enables precise measurements of the source parameters over 12 billion years of cosmic time.  %Intrinsic  masses will be determined with errors of $0.1\%-1\%$ for both MBHs, and the two spins with an absolute uncertainty as small as 0.01, in the best case. 
  %The most stringent requirement is set by being able to measure the spin of a \tb{threshold system [explain in what sense]} with $M_{\rm B}\sim 10^5\msun$, mass ratio $q\sim 0.1$ at $z\sim 3$. Higher masses, mass ratios and spins will result in louder signals and better parameter estimations. 
  %With luminosity distances determined to better than 10\% 
%and sky localization to better than one deg$^2$, we will learn whether the GW signal might be accompanied by detectable EM emission.
MBH coalescences may not occur in vacuum, and low-redshift ($z\lesssim 2$) binaries of $10^5<M_{\rm B}/\msun<10^7$  surrounded by circumbinary gas, may outshine in the optical and X-rays during the inspiral and merger proper, becoming key targets for EM follow up, with advance warning of hours. 
These mergers will be localized within $10$ or even $0.4 \,\rm deg^2$, corresponding to the field of view of Large Synoptic Survey Telescope and of the Athena  WFI \citep{McGee2018}, respectively.  The science with contemporaneous EM and GW observations is spectacular. It has the potential to discover the yet unknown periodic emission from shocked gas surrounding the two MBHs in the violently changing spacetime before merger \citep{Armitage02,Haiman17,Tang2018,Bowen2018,Ascoli18}, flashes, bursts and jetted emission at merger \citep{Palenzuela10,Kelly17}, and also post-merger afterglow signatures \cite{Rossi2010}. Linking masses and spins determined with exquisite precision by the GW signal 
with EM emission will be paramount.

\bigskip

\noindent {\Large \bf Deciphering the Astrophysics Behind the Discovery}

\medskip

%A MBH is some 12 orders of magnitude smaller than its host galaxy \tb{[Should clarify that this means that MBH is more spatially compact, not less massive]}.
At all redshifts, forming a MBH binary after a galaxy merger requires dissipation of orbital energy and efficient transport of angular momentum from the galaxy scale of hundreds of thousands of parsecs to the micro-parsec scale, when the merger gives birth to a new, single MBH \citep{bbr80,khan11,khan13,vasiliev14,Colpi14,Holley2015,Khan16,Souza17}. 
The physics governing the orbital evolution of a MBH pair on each scale is dramatically different. The process starts with the assembly of galaxy haloes and is followed by galaxy collisions, which all occur on cosmological scales \citep{tremmel18}. In the new galaxy, the pairing, hardening, and coalescence of two MBHs is a complicated dynamical problem. The processes of galaxy  and MBH binary dynamics are intimately connected. The link is provided by several strands of highly-coupled non-linear physics that ignites star formation, triggers nuclear inflows of gas, excites stellar and AGN feedback, and transports the incoming MBHs toward the center of the newly formed galaxy host. The cover page here depicts the merger of two galaxies and their embedded MBHs (circles), extracted from the cosmological simulation Romulus25, which tracks MBH pairs down to sub-kpc distances \citep{tremmel17, tremmel18}, still too widely separated for GWs to dominate, yet at the frontier of cosmological simulations of our day.  Given the overwhelmingly large dynamical range involved in this problem, numerical simulations coupled to semi-analytical models and sub-grid physics are precious tools to assist us in the interpretation of LISA data. Masses, mass ratios, eccentricities and spins, which are encoded in the GW signal, can be connected to the  physical processes leading to MBH binary formation and growth. For example, the eccentricity is largely amplified by stellar scatterings \citep{sesanaecc10,mirza17}, which is determined by the shape and kinematics of the background stellar potential. Meanwhile, the mass ratios and encounter geometries determine the efficiency of dynamical friction and sinking times \citep{Callegari11}, and spin magnitudes and orientations reflect BH interactions with massive gas discs \citep{BertiVolonteri08}. {\it Therefore, not only will LISA detect MBH binaries at the very end of their journey, but it will also unveil the cosmic evolution of the interplay between MBH binary dynamics and their host galaxies properties as they co-assemble in the cosmic web.}

%There is however a tangible chance that none of these searches will provide definitive detections of individual MBHs coalescences in the next decade. 

%There are about $10^{11}$ galaxies in the observable universe, each of which undergoes on average a merger in its lifetime. This gives about $10^{11}$ mergers in an Hubble time, or a rate of about 10 mergers per year. Detailed semianalytic models bracket this rate between few and few hundreds per year [REF?]. Information needs therefore to be extracted from precise measurements (masses, spins, redshifts) of a limited number of objects. 

The rates (3-20 per year in a conservative scenario) and properties of merging MBH binaries are inevitably connected with those of their host galaxies, and ultimately to the evolving large scale structure of the Universe \citep{Bonetti2018}. Complimentary to LISA, the
North American Nanohertz Observatory for GWs \citep{Nanograv18} and other
PTAs are targeting the GW foreground from very massive MBH binaries of $10^8-10^9 \msun$ at nHz frequencies observed during their inspiral phases up to $z\sim 1$ \citep{Chen17}. The spectrum of the GW foreground contains precious information on how the giant MBHs pair and interact with the broader galaxy dynamics.
Deciphering the information encoded in the LISA and PTA observations will grant us access to physics spanning a remarkable range, from the galactic scale down to the MBH horizon some 12 orders of magnitude smaller.  
In the coming years, observations of galaxies in deep fields coupled to forefront cosmological simulations will help us to interpret rate of MBH mergers as measured by LISA and PTAs in the low-frequency gravitational Universe. With its unique and nearly complete census of coalescing massive black hole binaries, from the Cosmic Dawn to the local Universe, GW observations will be a game changer in our understanding of the deepest mysteries of MBH birth, growth and coevolution, shedding light on structure formation, galaxy evolution and dynamics, accretion and fundamental physics.
% * <monica.colpi@mib.infn.it> 2018-12-19T09:09:36.184Z:
%
% ^.

%\noindent\fcolorbox{black}{pastelyellow}{%
%    \minipage[t]{\dimexpr1.00\linewidth-2\fboxsep-2\fboxrule\relax}
%%\begin{center} {\large\bf Summary} \end{center}
%{

%%%%%%%%%%%%%%%%%%%%%%%%%%%%%%%%%%%%%%%%%%%%%
%% Box 3
%%%%%%%%%%%%%%%%%%%%%%%%%%%%%%%%%%%%%%%%%%%%%
\bigskip

%\noindent \begin{minipage}{9cm}
% \begin{flushleft}
\noindent \begin{tcolorbox}[colback=pastelyellow,%gray background
                  colframe=black,% black frame colour
                  width=17cm,% Use 5cm total width,
                  arc=3mm, auto outer arc,
                 ]
 
%\end{tcolorbox}
%\noindent\fcolorbox{black}{pastelyellow}{%
%    \minipage[!t]{\dimexpr1.00\linewidth-2\fboxsep-2\fboxrule\relax}
    \smallskip
Space-based and pulsar timing gravitational wave observatories will cement the role of GWs as precise MBH probes across cosmic history, providing definitive answers about their origins and evolution. Interpreting the GW view of MBHs in the context of large scale structure, galaxy formation, and evolution requires a broad scientific vision that includes detailed modeling, inference, statistics, and input from EM surveys. By using MBH mergers as signposts for galaxy formation and assembly, we are poised for a paradigm shift in our understanding of MBHs and the Universe.
 \end{tcolorbox}

%the role of merger, accretion and feedback in shaping the properties of the massive BHs of our universe and their host galaxies,  which all impact on our view of the cosmos, solving clues and posing new questions. The impact of LISA's discoveries will be groundbreaking.

%}
%LISA will provide the first census to learn about their formation and cosmic evolution in concordance with the evolution of their host galaxies.
%LISA will be a game changer in our understanding of structure formation, galaxy evolution and dynamics, accretion and fundamental physics.
%What is the massive black hole coalescence rate and how is it related to the galaxy merger rate? 
%What can the spin of coalescing massive binary black holes tell us about their evolution ruled by mergers and accretion? 
%     \endminipage}\hfill\\ 

\newpage
\bibliographystyle{unsrtnat}
\bibliography{tb_smbh}
\bigskip
$^1$ M. Colpi, Department of Physics, University of Milano Bicocca, Piazza della Scienza 3, I20126 Milano, Italy

\smallskip
\noindent
$^2$ K. Holley-Bockelmann, Physics and Astronomy Department, Vanderbilt University, PMB 401807, 2301 Vanderbilt Place, Nashville, USA

\smallskip
\noindent
$^3$ K. Holley-Bockelmann, Physics Department, Fisk University, Nashville, USA

\smallskip
\noindent
$^4$ T.Bogdanovi\'c, Center for Relativistic Astrophysics, School of Physics, Georgia Institute of Technology, Atlanta GA 30332, USA

\smallskip
\noindent
$^5$ P. Natarajan, Department of Astronomy, Yale University, New Haven, CT 06511, USA

\smallskip
\noindent
$^6$ J. Bellovary, Queensboro Community College and American Museum of Natural History, New York, NY, USA

\smallskip
\noindent
$^7$ A. Sesana, School of Physics and Astronomy, University of Birmingham, Edgbaston, Birmingham B15 2TT, United Kingdom

\smallskip
\noindent
$^8$ M. Tremmel, Yale Center for Astronomy  and  Astrophysics, Physics Department, P.O. Box 208120, New Haven, CT 06520, USA

\smallskip
\noindent
$^9$ J. Schnittman, NASA Goddard Space Flight Center, Greenbelt, MD, USA

\smallskip
\noindent
$^{10}$ J. M. Comerford, Department of Astrophysical and Planetary Sciences, University of Colorado Boulder, Boulder, CO 80309, USA

\smallskip
\noindent
$^{11}$ E. Barausse, CNRS, UMR 7095, Institut d’Astrophysique de Paris, 98 bis Bd Arago, 75014 Paris, France

\smallskip
\noindent
$^{12}$ E. Berti, Department of Physics and Astronomy, Johns Hopkins University, 3400 N. Charles Street, Baltimore, MD 21218, USA

\smallskip
\noindent
$^{13}$ M. Volonteri, Sorbonne Universit\'es, UPMC Universit\'e Paris 06 et CNRS, UMR7095, Institut d’Astrophysique de Paris, 98bis Boulevard Arago, F-75014, Paris, France

\smallskip
\noindent
$^{14}$ F. M. Khan, Department of Space Science, Institute of Space Technology, P.O. Box 2750 Islamabad, Pakistan

\smallskip
\noindent
$^{15}$ S. T. McWilliams, Department of Physics and Astronomy, West Virginia University, Morgantown, WV 26506, USA; Center for Gravitational Waves and Astronomy, Morgantown, WV 26506, USA

\smallskip
\noindent
$^{16}$ S. Burke-Spolaor, Department of Physics and Astronomy, West Virginia University, Morgantown, WV 26506, USA; Center for Gravitational Waves and Astronomy, Morgantown, WV 26506, USA; CIFAR Azrieli Global Scholar

\smallskip
\noindent
$^{17}$ J. S. Hazboun, Physical Sciences Division, University of Washington Bothell, 18115 Campus Way NE Bothell, WA 98011-8246

\smallskip
\noindent
$^{18}$ J. Conklin, University of Florida

\smallskip
\noindent
$^{19}$ G. Mueller, University of Florida

\smallskip
\noindent
$^{20}$ S. Larson, Northwestern University

\end{document}